\documentstyle[12pt,titlepage]{article}

\font\grande=cmr9.5 scaled \magstep4
\font\medio=cmr9.5 scaled \magstep2
\outer\def\beginsection#1\par{\medbreak\bigskip
      \message{#1}\leftline{\bf#1}\nobreak\medskip
\vskip-\parskip
      \noindent}
\def\laq{\raise 0.4ex\hbox{$<$}\kern -0.8em\lower 0.62
ex\hbox{$\sim$}}
\def\gaq{\raise 0.4ex\hbox{$>$}\kern -0.7em\lower 0.62
ex\hbox{$\sim$}}

\begin{document}
\bibliographystyle {unsrt}

\titlepage

\begin{flushright}
CERN-PH-TH/2012-027
\end{flushright}

\vspace{15mm}
\begin{center}
{\grande Weyl invariance and the conductivity}\\
\vspace{5mm}
{\grande of the protoinflationary plasma}\\
\vspace{15mm}
 Massimo Giovannini 
 \footnote{Electronic address: massimo.giovannini@cern.ch} \\
\vspace{0.5cm}
{{\sl Department of Physics, Theory Division, CERN, 1211 Geneva 23, Switzerland }}\\
\vspace{1cm}
{{\sl INFN, Section of Milan-Bicocca, 20126 Milan, Italy}}
\vspace*{2cm}

\end{center}

\vskip 1.5cm
\centerline{\medio  Abstract}
We consider a globally neutral Lorentzian plasma as 
a possible remnant of a preinflationary stage of expansion and pose the problem of the suitable 
initial conditions for the evolution of the large-scale electromagnetic inhomogeneities. 
During the protoinflationary regime the  Weyl invariance of the Ohmic current guarantees 
that the comoving conductivity is approximately 
constant. The subsequent breaking of Weyl invariance by the masses of the charge carriers drives 
the conductivity to zero. The newly derived conducting initial conditions for the amplification of large-scale 
magnetic fields are contrasted with the conventional vacuum initial conditions. It is shown, in a specific 
class of examples, that when the number of inflationary efolds is close to minimal the effects of the conducting 
initial conditions cannot be neglected. 
\noindent

\vspace{5mm}

\vfill
\newpage
In recent years attention has been paid to the generation of large-scale magnetic fields in the early Universe \cite{rev}. Of particular interest for the present analysis are those mechanisms and toy models whose purpose is to amplify the vacuum fluctuations of the electromagnetic fields potentially present at the onset of the inflationary phase or, more accurately, around $63$ efolds prior to the end of inflation. Inflation cannot be however eternal in its past and the inflationary stage of expansion is normally complemented by a pre-inflationary stage where the evolution of the background is decelerated and, most likely, dominated by radiation \cite{nonvac}.  Protoinflationary initial conditions different from the vacuum have been discussed  by various authors \cite{nonvac} in connection with the scalar and tensor modes of the geometry and it is 
both interesting and legitimate to scrutinize the same situation in the case of large-scale electromagnetic fields. The purpose of the present 
study is to relax and complement the conventional approaches based on vacuum initial conditions 
where the length of the inflationary phase is assumed to be,  {\em a priori}, largely immaterial for the physical and mathematical aspects of the problem. The question we ought to address is, in short, the following: which are the initial conditions to be imposed on the evolution of the electromagnetic fields when, prior to the onset of inflation, a globally neutral plasma was present?  In the case of the scalar and tensor modes of the geometry it suffices to contemplate, for some applications, the possibility of a non-vanishing number of phonons (or gravitons) in the initial state. In the electromagnetic case the situation is physically different since the electric and magnetic fields do not interact in the same way with the plasma. 

Consider a conformally flat Friedmann-Robertson-Walker (FRW) metric $g_{\mu\nu}$ characterized by the scale factor $a(t)$. The evolution equations of the Hubble rate $H = \dot{a}/a$ are
\begin{equation}
H^2 M_{\mathrm{P}}^2 = \frac{8\pi}{3} \biggl[ \overline{\rho}_{\mathrm{tot}} + \frac{\dot{\varphi}^2}{2} + V(\varphi)\biggr] ,\qquad 
\dot{H} M_{\mathrm{P}}^2 =- 4\pi \biggl[ \dot{\varphi}^2 + (\overline{\rho}_{\mathrm{tot}} + \overline{P}_{\mathrm{tot}}) \biggr], 
\label{ff1}
\end{equation}
where $\varphi$ denotes the inflaton field with potential $V(\varphi)$ and the overdot denotes a derivation with respect to 
the cosmic time coordinate $t$; $\overline{\rho}_{\mathrm{tot}}$ and $\overline{P}_{\mathrm{tot}}$ are the 
covariantly conserved energy density and pressure of the plasma (i.e. $\partial_{t} \overline{\rho}_{\mathrm{tot}} + 3 H (\overline{\rho}_{\mathrm{tot}} + \overline{P}_{\mathrm{tot}}) =0$). The conformally 
flat metric $g_{\mu\nu}$ can be explicitly written as
$g_{\mu\nu}(\tau) = a^2(\tau)\eta_{\mu\nu}$ where $\eta_{\mu\nu}$ is the Minkowski metric and the conformal time coordinate $\tau$ is related to the cosmic time $t$ as $a(\tau) d\tau = d t$.  In the case $\overline{\rho}_{\mathrm{tot}} = 3 \overline{P}_{\mathrm{tot}}$ we shall fix the attention on solutions of Eq. (\ref{ff1}) with the following property  
\begin{equation}
\lim_{H_{1 } t \ll 1} a(t) \simeq (H_{1}\, t)^{1/2}, \qquad \delta \simeq \frac{\rho_{1}}{3 H_{1}^2 M_{\mathrm{P}}^2},
\label{ff3}
\end{equation}
where the parameter $\delta<1$ measures the relative weight of the radiation background during the 
protoinflationary stage when the slow-roll parameter $\epsilon = - \dot{H}/H^2$ is just about to drop below $1$ and the second time derivative of the scale factor (i.e. $\ddot{a}$) is turning from negative to positive. 
In Eq. (\ref{ff1}) and hereunder the physical variables (as opposed to their comoving counterparts) are denoted by a bar. The preinflationary (and hence decelerated) stage of expansion occurs for $t \ll  H_{1}^{-1}$. The protoinflationary regime corresponds instead to $t \simeq {\mathcal O}(H_{1}^{-1})$; finally, for $t \gg H_{1}^{-1}$ the Universe inflates. 

The total energy and pressure appearing in Eq. (\ref{ff1}) depend on the charged and neutral species. The contribution of the charged species to the transport coefficients is computed under the hypothesis that the collisions between the particles of the same charge can be neglected (i.e. $\Gamma_{\pm} \gg \Gamma_{+}$ and $\Gamma_{\pm} \gg \Gamma_{-}$), as it happens for Lorentzian plasmas \cite{KT}. Furthermore, at high temperatures $\Gamma_{\pm} > H$  for interactions mediated by massless gauge bosons.  Denoting with $m_{\pm}$ the masses of either the positive or negative charge carriers, their corresponding temperatures are the same as long as $\overline{T}_{\pm} \gg m_{\pm}$. When the species become non-relativistic (i.e. $\overline{T}_{\pm} < m_{\pm}$) the evolution
of the temperature depends on the relative hierarchy of the charged and neutral concentrations. In the latter case 
the entropy conservation and the first principle of thermodynamics imply: 
\begin{equation}
d[ a^2 ( \overline{T}_{+} + \overline{T}_{-})] + \gamma \, a \, d ( a \overline{T}_{r}) =0, \qquad \gamma = \frac{4\pi^4}{45 \zeta(3)}\frac{n_{r}}{n_{0}}, \qquad n_{+} = n_{-} = n_{0},
\label{pp1}
\end{equation}
where, following the conventions of Eq. (\ref{ff1}), $n_{\pm}= a^{3} \overline{n}_{\pm}$ and $n_{r} =a^3 \overline{n}_{r}$ denote the comoving concentrations. During the preinflationary and the protoinflationary stages of expansion (see Eq. (\ref{ff3}))  the radiation background dominates $\overline{\rho}_{\mathrm{tot}}$ (i.e. 
$n_{\mathrm{r}} \gg n_{0}$ and $\gamma\gg 1$). Equation (\ref{pp1}) can be solved, to lowest 
order, for $\overline{T}_{+} \simeq \overline{T}_{-} \simeq \overline{T}_{r}$ and the three temperatures 
scale, approximately, as $\overline{T} \sim a^{-\kappa }$ where $\kappa = ( 4 + \gamma)/(2 + \gamma) \simeq 1 + 2/\gamma + {\mathcal O}(1/\gamma^2)$. 
Note that $\gamma \propto g_{\mathrm{plasma}}^{-2}$  where $g_{\mathrm{plasma}} = 1/(n_{0} \lambda_{\mathrm{D}}^3)\ll 1$ is the plasma parameter
quantifying, by definition, the inverse of the number of particles of charge $q$ present in the Debye sphere (i.e. the sphere whose radius is given by the Debye length $\lambda_{\mathrm{D}} = \sqrt{T/(8\pi\, n_{0} q^2)}$ ).
Provided $g_{\mathrm{plasma}} \ll 1$ (i.e. when the Debye sphere contains a large number of particles) $\gamma$ is 
also much larger than $1$; in a related context this requirement goes under the name of plasma approximation \cite{PB}.

The contribution of the charged species to the Maxwell equations remains
unsuppressed in the protoinflationary phase and, depending on the total number of efolds, during the early stages of the inflationary epoch. Indeed, for $\overline{T} \gg m_{\pm}$ (i.e. $T \gg m_{\pm} a$) the whole evolution of the charged species is Weyl invariant and this symmetry prevents the suppression of the conductivity in the relativistic regime. 
The simplest way to demonstrate the Weyl invariance of the full 
system rests on the explicit expression of the Vlasov-Landau equations 
for the distribution functions of the charged species written in a conformally flat 
metric of FRW type (see, e.g. \cite{KT}):
\begin{equation}
\frac{\partial f_{\pm}}{\partial \tau} + \vec{v}_{\pm} \cdot \vec{\nabla}_{\vec{x}} f_{\pm}  \mp q \biggl[ \vec{E} 
+ \vec{v}_{\pm} \times \vec{B} \biggr] \cdot \vec{\nabla}_{\vec{p}} f_{\pm} = \biggl(\frac{\partial f_{\pm}}{\partial \tau} \biggr)_{\mathrm{coll}},
\label{ff5}
\end{equation}
where $\vec{E} = a^2 \vec{{\mathcal E}}$, $\vec{B} = a^2 \vec{{\mathcal B}}$  and $\vec{v}_{\pm} = \vec{p}/\sqrt{p^2 + m_{\pm}^2 a^2 }$; $\vec{p}$ denotes the comoving three-momentum. When the mass contribution can be neglected in comparison with $p^2$ Eq. (\ref{ff5}) has the same form it would have in Minkowski space-time \cite{KT}. 
In the opposite situation (i.e.  $p^2\ll m_{\pm}^2 a^2$) the mass term breaks explicitly Weyl invariance. The comoving concentrations of the charged species can be obtained by integrating the distribution functions over the comoving three-momenta so that the Maxwell equations are:
\begin{eqnarray}
&& \vec{\nabla}\cdot \vec{E} = 4 \pi q (n_{+} - n_{-}), \qquad \vec{\nabla}\cdot \vec{B} =0,
\label{ff7}\\
&& \vec{\nabla} \times \vec{E} = - \frac{\partial \vec{B}}{\partial\tau}, \qquad \vec{\nabla} \times \vec{B} = 4 \pi q (n_{+} 
\vec{v}_{+} - n_{-} \vec{v}_{-}) + \frac{\partial \vec{E}}{\partial \tau}.
\label{ff8}
\end{eqnarray}
Eqs. (\ref{ff5}), (\ref{ff7}) and (\ref{ff8}) are then Weyl invariant in the relativistic regime and in the plasma approximation 
which guarantees an approximate common temperature of the charged and neutral species (see Eq. (\ref{pp1})). The latter result is quite known\footnote{This conclusion can be found within different perspectives in the book and in the papers reported in \cite{lic}.} and can be directly obtained, 
 by means of simple scaling considerations, from the generally covariant form of Maxwell equations, i.e.\footnote{Note that  $f_{\mu\nu}$ and $\tilde{f}_{\mu\nu}$ are, respectively, the Maxwell field strength and its dual; $g$ denotes, as usual,  the determinant of $g_{\mu\nu}$; in terms of the physical fields ${\mathcal E}^{i}$ and ${\mathcal B}^{i}$ 
we have $f^{i 0} = {\mathcal E}^{i}/a^2$ and $f^{ij} = - \epsilon^{i j k} \,\,{\mathcal B}_{k}/a^2$. Furthermore, as already 
mentioned after Eq. (\ref{ff5}), comoving and physical fields are related
 as $\vec{E} = a^2 \vec{{\mathcal E}}$ and $\vec{B} = a^2 \vec{{\mathcal B}}$.}:
  \begin{equation}
  \partial_{\mu} \bigl( \sqrt{-g} \, f^{\mu\nu} \bigr) = 4 \pi\, \sqrt{- g} \,j^{\nu},\qquad    \partial_{\mu} \bigl( \sqrt{-g} \, \tilde{f}^{\mu\nu} \bigr) = 0.
  \label{sim1}
  \end{equation}
In four space-time dimensions 
both $\sqrt{-g}  f^{\mu\nu}$ and $\sqrt{-g}  \tilde{f}^{\mu\nu}$ are  invariant under the Weyl rescaling 
 of the form $g_{\mu\nu}(x) = a^2(x) \eta_{\mu\nu}$ where $x$ labels a generic space-time coordinate. Consequently the whole 
system of Eq. (\ref{sim1}) is Weyl invariant if $\sqrt{-g} j^{\nu}$ is separately Weyl invariant.  Consider, in this respect,  a relativistic plasma with Ohmic current $j^{\nu} = \overline{\sigma} f^{\mu\nu}\,\overline{u}_{\mu}$ where $\overline{\sigma}(x)$ is the physical conductivity.  Since $\sqrt{-g}  f^{\mu\nu}$ is separately Weyl invariant, the 
expression  $\sqrt{-g} j^{\nu}$ is Weyl invariant provided the combination  $\overline{\sigma} \,\overline{u}_{\mu}$ 
is invariant. Under Weyl rescaling $\overline{u}_{\mu}$ transforms as 
$\overline{u}_{\mu} \to u_{\mu}= \overline{u}_{\mu}/a(x)$ because of $g^{\mu\nu} \overline{u}_{\mu} \overline{u}_{\nu} =1$.
For this reason $\sqrt{-g} j^{\nu}$ is Weyl invariant provided
the conductivity scales as $\overline{\sigma}(x) \to \sigma(x) = \overline{\sigma}(x) a(x)$.  But this happens exactly in a relativistic plasma, where the physical conductivity scales as the first power of the physical temperature (i.e. $\overline{\sigma}(x) \propto \overline{T}(x)$) and it coincides, in practice, with the comoving conductivity $\sigma(x)  \propto T(x) = \overline{T}(x) a(x)$ which is approximately constant whenever $\overline{T} \sim a^{-1}$. 

The lack of Weyl invariance of Eqs. (\ref{ff5})--(\ref{ff8}) when $\overline{T} < m_{\pm}$ is reflected in the general form of the conductivity;  to simplify the expressions a hierarchy in the masses of the  charge carriers can be assumed (for instance $m_{+} > m_{-} = m$) so that, for a Lorentzian plasma \cite{KT}, the conductivity is: 
\begin{equation}
\sigma(a,\gamma) = \frac{T(a,\gamma)}{q^2 \sqrt{1 + \frac{m a}{T(a,\gamma)}}},\qquad \lim_{\gamma \gg 1} T(a,\gamma) \propto a^{- 2/(\gamma +2)} = \mathrm{constant}.
\label{ff9}
\end{equation}
In the limit $\gamma \gg 1$  the comoving temperature $ T = \overline{T} a$ is approximately constant; different situations, corresponding to specific values of $\gamma$, can be studied but they are less significant for the illustrative purposes of the present analysis.  In the limit $T \gg m a$ Eq. (\ref{ff9}) implies $\sigma \simeq T/q^2$ (as it happens in the case of an relativistic plasma); in the opposite limit, $\sigma \simeq T/q^2 \sqrt{T/(m a)}$. By subtracting the evolution equations of the velocities according to the standard procedure \cite{PB}, it is possible to obtain the equation for the total current appearing in Eq. (\ref{ff8}): 
\begin{eqnarray}
\frac{\partial \vec{J}}{\partial \tau} + a( H + \Gamma_{\pm} ) \vec{J} = \frac{\omega_{\mathrm{p}}^2 }{4\pi} (\vec{E} + \vec{v}\times \vec{B} )\qquad \vec{J} = q ( n_{+} \vec{v}_{+} - n_{-} \vec{v}_{-}),
\label{ff10}
\end{eqnarray}
where terms like $\vec{J}\times \vec{B}$ have been neglected and where $\vec{v} = (m_{+} \vec{v}_{+} + m_{-} \vec{v}_{-})/(m_{+} + m_{-})$. Since $\Gamma_{\pm} > H$ the total Ohmic current becomes indeed 
$\vec{J} \simeq \sigma (\vec{E} + \vec{v}\times \vec{B})$ with $\sigma = \omega_{\mathrm{p}}^2/(4 \pi a \Gamma_{\pm})$.

Equation (\ref{ff9}) can be phrased in terms of the number of inflationary efolds $N = \ln{(a/a_{\mathrm{1}})}$ 
where $a_{\mathrm{1}}$ denotes the scale factor at the onset of the inflationary phase; thus we have
$\sigma(N)= \sigma_{*}/\sqrt{1 + e^{N - N_{*}}}$. For $N < N_{*}$ the conductivity is approximately constant while for $N> N_{*}$ it is exponentially suppressed. To leading order in $g_{\mathrm{plasma}}$ the critical number of efolds $N_{*}$  depends on the temperature reached during the protoinflationary phase and it can be estimated by recalling that at the onset of the inflationary phase $\delta <1$ in Eq. (\ref{ff3}). Requiring that, at most, $\delta \simeq 1$ 
an upper bound on $N_{*}$ can be obtained:
\begin{equation}
N_{*} = -0.253+ \frac{1}{2} \ln{\xi} - \frac{1}{4} \ln{g_{\mathrm{th}}} - \ln{\biggl(\frac{m}{M_{\mathrm{P}}}\biggr)},
\label{ff11}
\end{equation}
where $g_{\mathrm{th}}$ denotes the effective number of relativistic degrees of freedom; $\xi =H_{\mathrm{1}}/M_{\mathrm{P}}= \sqrt{\pi \epsilon {\mathcal A}_{{\mathcal R}}}$ where
 ${\mathcal A}_{\mathcal R} = 2.43\times 10^{-9}$ is the amplitude of the scalar power 
spectrum at the pivot scale $k_{\mathrm{p}} =0.002\, \mathrm{Mpc}^{-1}$ and $\epsilon$ is the slow-roll 
parameter introduced after Eq. (\ref{ff3}). For the numerical estimates we shall adopt the values of the cosmological parameters determined in terms of the WMAP data alone in the minimal concordance model (see last three papers of Ref. \cite{WMAP7} for the latest release). Since the maximal protoinflationary temperature should not exceed the energy density of the background, we have that, at most, $N_{*} \simeq 36.78 - 0.25 \ln{(g_{\mathrm{th}}/100)} + 0.5 \ln{(\xi/10^{-5})} - \ln{(m/\mathrm{GeV})}$.
The value of $N_{*}$ can be compared with the maximal number of efolds presently accessible by large-scale observations, i.e.
\begin{equation}
N_{1}  = 62.2 + \frac{1}{2} \ln{\biggl(\frac{\xi}{10^{-5}}\biggr)} - \ln{\biggl(\frac{h_{0}}{0.7}\biggr)}
+ \frac{1}{4} \ln{\biggl(\frac{h_{0}^2 \, \Omega_{\mathrm{R}0}}{4.15\times 10^{-5}}\biggr)},
\label{DS25B}
\end{equation}
which is close, by construction, to the minimal number of efolds $N_{\mathrm{min}}$ needed to solve the kinematic 
problems of the standard cosmological model (i.e. $N_{\mathrm{min}} \simeq N_{1}$). Recalling the fiducial set of 
cosmological parameters quoted before \cite{WMAP7}, Eq. (\ref{DS25B}) gives $N_{1} = 63.6 + 0.25 \ln{\epsilon}$.
Equation (\ref{DS25B}) is derived in the sudden reheating approximation but $N_{1}$ can be larger if right after inflation the Universe expands at a rate which is slower than radiation  down to the big-bang nucleosynthesis curvature scale. In the latter case  the estimate for $N_{1}$ increases by ${\mathcal O}(14)$ efolds so that  $N_{1} \simeq 78.3 + 0.33 \ln{\epsilon}$ in agreement with previous estimates \cite{SR}.

If  $N \simeq N_{\mathrm{tot}} \gg N_{*} + N_{\mathrm{min}}$, Weyl invariance is broken before the onset of the last $N_{1} \sim {\mathcal O}(63)$ efolds of inflationary expansion; the electromagnetic fields are normalized when the 
the protoinflationary conductivity is suppressed as $ e^{-(N_{\mathrm{tot}} - N_{*})/2}$. But this 
means that the sources do not contribute, in practice, to the initial conditions which are accurately fixed by quantum mechanics. 
Conversely if $N_{\mathrm{min}}< N \leq N_{*} + N_{\mathrm{min}}$ the conductivity will be constant for the first $N_{*}$ efolds 
and then it will be exponentially suppressed as 
$e^{(N_{*} -N_{\mathrm{min}})/2}$ (if $N\sim N_{\mathrm{min}}$) and as $e^{-N_{\mathrm{min}}/2}$ (if $N\sim N_{\mathrm{min}} + N_{*}$). While the point 
here is not to endorse a specific duration of the inflationary phase, when $N_{\mathrm{min}}< N \leq N_{*} + N_{\mathrm{min}}$ the last ${\mathcal O}(63)$ efolds of inflationary expansion may start when the conductivity did not undergo a substantial  suppression. For instance, when the mass range of the lightest charge carrier is ${\mathcal O}(\mathrm{GeV})$ and if $N\sim {\mathcal O}(N_{\mathrm{min}})$ the conductivity is still almost constant ${\mathcal O}(30)$ efolds prior to the end of inflation. In this class of physical situations the normalization of the electric and magnetic fields does not follow from the quantum mechanical initial conditions but rather from the standard conducting initial conditions \cite{PB}.  From Eqs. (\ref{ff7})--(\ref{ff8}) and (\ref{ff10}) $\vec{E}$ is solenoidal (because of the global charge 
neutrality) but also $\vec{J}$ must be solenoidal since in the plasma rest frame $\vec{E} \sim \vec{J}/\sigma$. 
In the regime of high conductivity the displacement current can be neglected and therefore the appropriate 
initial conditions for the electromagnetic fields at $\tau_{*}$ are simply given by 
\begin{equation}
\vec{B}(\vec{x},\tau_{*}) = \vec{B}^{(\mathrm{in})}(\vec{x}), \qquad \vec{E}(\vec{x},\tau_{*}) = \frac{\vec{\nabla} \times
\vec{B}^{(\mathrm{in})}(\vec{x})}{4 \pi \sigma_{*}},
\label{sol2}
\end{equation}
together with the conditions $\vec{\nabla}\cdot\vec{E} = \vec{\nabla} \cdot \vec{B} = \vec{\nabla}\cdot \vec{J} =0$. 
Let us now characterize, for immediate convenience, the electric and magnetic fields by means of their associated power spectra defined, in Fourier space, as 
\begin{equation}
\langle B_{i}(\vec{k},\tau) \, B_{j}(\vec{p},\tau) \rangle = \frac{2 \pi^2}{k^3} P_{B}(k,\tau) P_{ij}(\hat{k}) \delta^{(3)}(\vec{k} + \vec{p}),
\label{sol2a}
\end{equation}
where $P_{ij}(\hat{k})= \delta_{ij} - \hat{k}_{i} \hat{k}_{j}$ is the transverse projector; exactly the same definition 
holds in the case of the electric fields whose related power spectrum will be denoted by $P_{\mathrm{E}}(k,\tau)$.

The vacuum and the conducting initial conditions lead to different power spectra which can be compared.
Consider, for instance, the class of models where the gauge coupling is dynamical during inflation
according to the action $\int d^{4} x \sqrt{- g} \sqrt{\lambda} f_{\mu\nu} f^{\mu\nu}/(16\pi)$ \cite{DT}. 
When $\lambda$ increases the gauge coupling decreases and this possibility fits with the presence of a curvature singularity in the past history of the inflationary dynamics: since the gravitational coupling gets strong, it is also appropriate to contemplate the case when $\sqrt{\lambda}$  is initially ${\mathcal O}(1)$, it increases  during the inflationary phase as $\sqrt{\lambda} \sim (- \tau)^{1/2 - \nu}$ and it decreases again during the reheating process (see last four articles of Ref. \cite{DT}).
The Hamiltonian describing the evolution of the classical electromagnetic inhomogeneities is given, in Fourier space, by
\begin{equation}
{\mathcal H}(\tau)  = \frac{1}{2} \int d^3 k \biggl[ \vec{\pi}_{\vec{k}} \cdot \vec{\pi}_{-\vec{k}} +  \frac{(\sqrt{\lambda})'}{\sqrt{\lambda}}
\biggl( \vec{\pi}_{\vec{k}} \cdot \vec{y}_{-\vec{k}}  +  \vec{\pi}_{-\vec{k}} \cdot \vec{y}_{\vec{k}}\biggr)
+k^2 \vec{y}_{\vec{k}} \cdot \vec{y}_{-\vec{k}}\biggr],
\label{sol4}
\end{equation}
where the prime denotes a derivation with respect to the conformal time coordinate $\tau$. 
In the Coulomb gauge the (comoving) electromagnetic fields $\vec{B} = \vec{\nabla}\times \vec{y}$ and  $\vec{E} = - \vec{\pi}$;  
Eq. (\ref{sol4}) is invariant under the duality transformation $\sqrt{\lambda} \to \frac{1}{\sqrt{\lambda}}$ and 
$ \vec{\pi}_{\vec{k}} \to - k \vec{y}_{\vec{k}}$, $\vec{y}_{\vec{k}} \to \frac{1}{k} \vec{\pi}_{\vec{k}}$ (see first two references in Ref. \cite{DT}). The Hamilton equations derived from Eq. (\ref{sol4}) become:
$\vec{y}_{\vec{k}}^{\,\prime} = \vec{\pi}_{\vec{k}} + [(\sqrt{\lambda})'/\sqrt{\lambda}] \vec{y}_{\vec{k}}$,
and $\vec{\pi}_{\vec{k}}^{\,\prime} = - k^2 \vec{y}_{\vec{k}} -[(\sqrt{\lambda})'/\sqrt{\lambda}]\vec{\pi}_{\vec{k}}$. 
In the case of the amplification of vacuum fluctuations the power spectra of the magnetic field have a spectral 
slope $k^{n_{\mathrm{B}}-1}$ with $n_{\mathrm{B}} = (6 - 2 \nu)$; in scale invariant limit (i.e. $\nu = 5/2$) 
the present value of the power spectrum can be estimated as \cite{DT}
\begin{equation}
\log{[\sqrt{P_{{\mathcal B}}(k,\tau_{0})}/\mathrm{Gauss}]} = -10.85 + 0.5 \biggl[\log{\biggl(\frac{{\mathcal A}_{{\mathcal R}}}{2.43 \times 10^{-9}}\biggr)} + \log{\biggl(\frac{\Omega_{\mathrm{R}0}}{4.15 \times 10^{-5}}\biggr)}\biggr],
\label{sol13}
\end{equation}
where $P_{\mathrm{B}}(k,\tau) = a^4(\tau) P_{{\mathcal B}}(k,\tau)$. The result of Eq. (\ref{sol13}) assumes, for consistency 
with Eqs. (\ref{ff11}) and (\ref{DS25B}), the sudden reheating approximation. When the modes of the field 
are inside the Hubble radius during inflation (i.e. $k\tau > 1$) the electric and the magnetic power 
spectra are of the same order (i.e. $P_{\mathrm{E}}(k,\tau) \simeq P_{\mathrm{B}}(k,\tau)$). In the opposite case (i.e. 
$k\tau \ll 1$) the electric power spectra are suppressed, in comparison with their magnetic counterpart, as $P_{\mathrm{E}}(k,\tau) \simeq |k\tau|^2 \,  P_{\mathrm{B}}(k,\tau)$. In the sudden reheating approximation the
conductivity raises again at the end of inflation and this step further suppresses exponentially the electric fields. 

The estimate of Eq. (\ref{sol13})  can be compared with magnetic field obtainable in the framework 
of the same toy model but in the case of the conducting initial conditions.
For sake of concreteness we will have that for $\tau < - \tau_{*}$ the conductivity and $\lambda$ will both be constant $\sigma = \sigma_{*}$ and $\sqrt{\lambda} = \sqrt{\lambda_{*}}$. For $\tau > - \tau_{*}$ the conductivity will vanish, $\partial_{\tau}\sqrt{\lambda} >0$ and we can assume, for sake of comparison with the vacuum case of Eq. (\ref{sol13}), that $\sqrt{\lambda} \simeq 
(-\tau)^{1/2 - \nu}$. Therefore, the electromagnetic fields subjected to the conducting initial 
conditions (\ref{sol2}) and subsequently amplifield by the evolution of $\sqrt{\lambda}$ are:
\begin{equation}
\vec{E}( k,\, \tau,\, \tau_{*}) = - i \hat{k} \times \vec{B}(k,\tau_{*})\, {\mathcal G}(x, z), \qquad \vec{B}(k, \tau, \tau_{*}) = 
\vec{B}(k,\tau_{*})\, {\mathcal F}(x, z),
\label{sol10}
\end{equation}
where $x = k \tau_{*}$ and $z= - k \tau$; recalling the definitions of Hankel functions of first and second kind (i.e. $H_{\nu}^{(1)}$ and $H_{\nu}^{(2)}$) it is useful to define the following combination $P_{\nu}^{(1)}(z) = 2 \nu H_{\nu}^{(1)}(z) - z H_{\nu + 1}^{(1)}(z)$ (and similarly for $P_{\nu}^{(2)}(z) = 2 \nu H_{\nu}^{(2)}(z) - z H_{\nu + 1}^{(2)}(z)$).  Consequently the functions ${\mathcal G}(x,z)$ and ${\mathcal F}(x,z)$ appearing in Eq. (\ref{sol10}) can be written as:
\begin{eqnarray}
{\mathcal G}(x,z) &=& \frac{i \pi}{4} \, \sqrt{\frac{x}{z}} \biggl\{ \frac{1}{x}\biggl[ P_{\nu}^{(1)}(z) P_{\nu}^{(2)}(x) 
- P_{\nu}^{(1)}(x) P_{\nu}^{(2)}(z)\biggr]  
\nonumber\\
&-& \eta \biggl[ H_{\nu}^{(2)}(x) P_{\nu}^{(1)}(z) - H_{\nu}^{(1)}(x) P_{\nu}^{(2)}(z)\biggr]\biggr\},
\nonumber\\
{\mathcal F}(x,z) &=& \frac{i \pi}{4} \, \sqrt{\frac{z}{x}}\biggl\{ \biggl( 2 \nu - \eta \, x \biggr) \biggl[H_{\nu}^{(2)}(x) 
H_{\nu}^{(1)}(z) - H_{\nu}^{(1)}(x) H_{\nu}^{(2)}(z) \biggr]
\nonumber\\
&+& x \biggl[ H_{\nu + 1}^{(1)}(x) H_{\nu}^{(2)}(z) - H_{\nu+1}^{(2)}(x) H_{\nu}^{(1)}(z)\biggr]\biggr\},
\label{sol11}
\end{eqnarray}
where $\eta = k/4\pi \sigma_{*}$. By setting $\tau= - \tau_{*}$ the expressions of Eq. (\ref{sol11}) reproduce the conducting initial conditions given in Eq. (\ref{sol2}).  Furthermore, the associated power spectra can be written, respectively,  as $P_{\mathrm{B}}(k,\tau,\tau_{*}) = P_{\mathrm{B}}(k,\tau_{*}) \, |{\mathcal F}(z, x)|^2$, and $P_{\mathrm{E}}(k,\tau,\tau_{*}) = P_{\mathrm{B}}(k,\tau_{*}) \, |{\mathcal G}(z, x)|^2$. In the limit 
$|\tau_{*}| \ll |\tau_{e}|$ (and $x <1$, $z<1$) the electric power spectra are always exponentially suppressed already during inflation in comparison  with the magnetic power spectra. From Eq. (\ref{sol11}), at the end of the 
inflationary phase, $P_{\mathrm{B}}(k,\tau_{e},\tau_{*}) \simeq P_{\mathrm{B}}(k,\tau_{*}) (a_{e}/a_{*})^{2\nu -1}$
and $P_{\mathrm{E}}(k,\tau_{e},\tau_{*}) \simeq \eta P_{\mathrm{B}}(k,\tau_{*}) \, (a_{e}/a_{*})^{1 - 2\nu}$.
Since $\eta = k/(4\pi \sigma)\simeq k/T_{*} \ll 1$ and $ 1/2 \leq \nu \leq 5/2$ the suppression 
of the electric fields is always much larger than in the case of vacuum initial conditions.
To assess the amplitude of the magnetic fields we shall consider, for sake of comparison with Eq. (\ref{sol13}) the case 
$\nu = 5/2$. To avoid supplementary assumptions, the amplitude of the magnetic power spectrum during the protoinflationary phase can be set to its maximum value compatible with the closure bounds. In the latter case the power spectrum of the magnetic field will be given by 
\begin{eqnarray}
\log{[\sqrt{P_{{\mathcal B}}(k,\tau_{*}, \tau_{0})}/\mathrm{Gauss}]} &=& -60.15 + 0.5\log{ \biggl(\frac{\Lambda}{10^{-4}}\biggr)}+ ( 2 \alpha -1) \log{\biggl(\frac{H_{r}}{H_{e}}\biggr)}
\nonumber\\
&+& 0.43[ (\nu - 1/2)(N - N_{*}) - 2 (N- N_1)],
\label{sol15}
\end{eqnarray}
where $\Lambda=  \sqrt{P_{{\mathcal B}}(k,\tau_{1})}/(H_{1}^2 M_{\mathrm{P}}^2) < 1$ measures the fraction of energy 
density stored in the magnetic field at $\tau_{1}$; $H_{r}$ accounts for the possibility of a delayed radiation-dominated phase between the end of inflation and the onset 
of big-bang nucleosynthesis. The exponent $\alpha$ depends on the expansion rate between the end 
of the inflationary phase and the onset of the standard (i.e. post-inflationary) radiation-dominated epoch. Equation (\ref{sol15}) has several interesting limits.
Recalling that the the interval of $\nu$ is restricted, from kinematical considerations, to $ 1/2 \leq \nu \leq 5/2$, in the case 
$\nu = 1/2$ there is no amplification due to the evolution of the gauge coupling and therefore the upper bound on the 
magnetic field intensity is around ${\mathcal O}(10^{-61})$ Gauss in the sudden reheating approximation 
where $H_{r} \sim H_{e}$. This is simply the magnetic field one would obtain from the 
protoinflationary initial conditions assuming the minimal amount of inflationary efolds.  Conversely, if $\nu = 5/2$ and $N = N_{1} \simeq {\mathcal O}(65)$ and $N_{*} \sim {\mathcal O}(35)$  the maximal magnetic field 
turns out to be  $10^{-35}$ Gauss (in the sudden reheating approximation) which can become of the order of ${\mathcal O}(10^{-23})$ Gauss for a stiff post-inflationary phase extending down to the nucleosynthesis scale.

The implications of a globally neutral plasma during the protoinflationary stage of expansion have been investigated.
This idea did not receive specific attention so far even if it seems rather natural in the light of similar attempts carried on in the 
case of the conventional scalar and tensor modes of the geometry. 
If the plasma is relativistic, Weyl invariance prevents the suppression of the conductivity which 
starts being diluted as soon as Weyl invariance gets broken by the masses of the charge carriers. 
The presence of a relativistic plasma in the initial conditions of the inflationary dynamics changes qualitatively and quantitatively the initial data to be imposed for the evolution of the large-scale electromagnetic inhomogeneities. 
The conducting initial conditions for the amplification of large-scale magnetic fields have been contrasted with the conventional vacuum initial conditions. In a class of specific examples the amplification of the magnetic fields and the suppression of the electric fields has been shown to depend explicitly on the number of efolds of the inflationary phase.


\begin{thebibliography}{99}

\bibitem{rev}  K. Enqvist, Int.\ J.\ Mod.\ Phys.\  D  {\bf 7}, 331 (1998);  M.~Giovannini,  Int.\ J.\ Mod.\ Phys.\  {\bf D13}, 391 (2004); 
J.~D.~Barrow, R.~Maartens and C.~G.~Tsagas,  Phys.\ Rept.\  {\bf 449}, 131 (2007).

\bibitem{nonvac} P.~D.~B.~Collins and R.~F.~Langbein,ÊÊPhys.\ Rev.\ D {\bf 45}, 3429 (1992); I.~Sokolov,ÊÊClass.\ Quant.\ Grav.\  {\bf 9}, L61 (1992);
 M.~Gasperini, M.~Giovannini, G.~Veneziano,  Phys.\ Rev.\  {\bf D48}, 439 (1993);
K.~Bhattacharya, S.~Mohanty and R.~Rangarajan, Phys.\ Rev.\ Lett.\  {\bf 96}, 121302 (2006);  W.~Zhao, D.~Baskaran and P.~Coles,
 ÊPhys.\ Lett.\ B {\bf 680}, 411 (2009); M.~Giovannini,  Phys.\ Rev.\ D {\bf 83}, 023515 (2011); I.~Agullo and L.~Parker,
ÊÊPhys.\ Rev.\ D {\bf 83}, 063526 (2011); R.~Lieu and T.~W.~B.~Kibble,
  arXiv:1110.1172 [astro-ph.CO]; S.~Kundu,  arXiv:1110.4688 [astro-ph.CO].

\bibitem{KT} E. M. Lifshitz and L. P. Pitaevskii, {\it Physical Kinetics} (Pergamon, Oxford, England, 1980); J. Bernstein, 
{\it Kinetic theory in the expanding universe} (Cambridge Univ. Press, Cambridge, England, 1988).

\bibitem{PB} N. A. Krall, A. W. Trivelpiece, {\it Principles of Plasma Physics}, (San Francisco Press, San Francisco 1986);
T.~J.~M Boyd and J.~J.~Sanderson {\it The Physics of Plasmas}, (Cambridge Univ. Press, Cambridge, UK, 2003).

\bibitem{lic} A. Lichnerowicz, {\it Magnetohydrodynamics: waves and shock waves in curved space-time}, (Kluwer academic publisher, 1994);
P. Olesen, Phys. Lett. B {\bf 398}, 321 (1997); A. Brandenburg, K. Enqvist and P. Olesen   Phys. \ Rev.\ D {\bf 54}, 1291  (1996);
 K.~Subramanian and J.~D.~Barrow, Phys.\ Rev.\ D {\bf 58}, 083502 (1998); M.~Christensson, M.~Hindmarsh,  Phys.\ Rev.\  {\bf D60}, 063001 (1999).
 
\bibitem{WMAP7}  D.~N.~Spergel {\it et al.},  Astrophys.\ J.\ Suppl.\  {\bf 148}, 175 (2003); D.~N.~Spergel {\it et al.},
{\em ibid.} \ {\bf 170}, 377 (2007);  C.~L.~Bennett {\it et al.}, {\em ibid.}\ {\bf 192}, 17 
(2011);  B.~Gold {\it et al.},  {\em ibid.} \ {\bf 192}, 15 (2011); 
E.~Komatsu {\it et al.},  {\em ibid.} \ {\bf 192}, 18 (2011).

\bibitem{SR}  M.~Giovannini,  Phys.\ Rev.\  {\bf D60}, 123511 (1999);
V.~Sahni, M.~Sami and T.~Souradeep, Phys.\ Rev.\  D {\bf 65}, 023518 (2002);  
A.~R. Liddle and S.~M. Leach,   Phys.\ Rev.\  {\bf D68}, 103503 (2003);
 H.~Tashiro, T.~Chiba and M.~Sasaki, Class.\ Quant.\ Grav.\  {\bf 21}, 1761 (2004);
 T.~J.~Battefeld and D.~A.~Easson,  Phys.\ Rev.\  D {\bf 70}, 103516 (2004).

\bibitem{DT}  S.~Deser and C.~Teitelboim,  Phys.\ Rev.\  D {\bf 13}, 1592 (1976); S. Deser,  
J. Phys. A {\bf 15},  1053 (1982);    B. Ratra,  Astrophys.\, J.\, Lett.  {\bf 391}, L1 (1992);  M. Gasperini,
 M. Giovannini, and G. Veneziano, Phys. Rev. Lett. {\bf 75}, 3796 (1995); M.~Giovannini,  Phys.\ Rev.\  D {\bf 64}, 061301 (2001); 
 K.~Bamba and M.~Sasaki,  JCAP {\bf 0702}, 030 (2007);  M. Giovannini, Phys.\ Lett.\  B {\bf 659}, 661 (2008);  JCAP {\bf 1004}, 003 (2010).


\end{thebibliography}
\end{document}